\documentstyle[aps,preprint]{revtex}
 
\tighten

\def\npb#1#2#3{{Nucl.~Phys.} {\bf B#1} (#2) #3}
\def\prd#1#2#3{{Phys.~Rev.} {\bf D#1} (#2) #3}
\def\plb#1#2#3{{Phys.~Lett.} {\bf #1B} (#2) #3}
\def\hth{hep-th}
\def\loe{low-energy}
\def\kpot{K\"ahler potential}

\def\suii{$SU(2)$}
\def\uone{$U(1)$}

\def\OO{{\cal O}}
\def\NN{{\cal N}}
\def\none{$\NN=1$}
\def\ntwo{$\NN=2$}
\def\nfour{$\NN=4$}
\def\susy{supersymmetry}
\def\susic{supersymmetric}

\def\hyp{hypermultiplet}
\def\slz{$SL(2,{\bf Z})$}
\def\tr{{\rm tr}}
\def\half{{1\over2}}

\def\bel{\begin{equation}\label}
\def\ee{\end{equation}}
\newcommand\eref[1]{Eq.~(\ref{#1})}
\newcommand\vev[1]{\langle{#1}\rangle}

\begin{document}
\pagestyle{empty}

\preprint{
\begin{minipage}[t]{3in}
\begin{flushright}
RU--96--109, IAS--TH--96/124, ILL--(TH)--96--16, 
CLNS--99/1642, NSF--ITP--99--126, UCSD/PTH 99-16
\\
hep-th/9910250
\\
October 1999
\end{flushright}
\end{minipage}
}

\title{On Inherited Duality\\
in $\NN=1$ $d=4$ Supersymmetric Gauge Theories} 

\author{
Philip C. Argyres
\thanks{Newman Laboratory, Cornell University, Ithaca NY 14853}
\thanks{Institute for Theoretical Physics, University of California, 
Santa Barbara CA 93106},
Ken Intriligator
\thanks{UCSD Physics Department, 9500 Gilman Drive, La Jolla CA 92093},\\
Robert G. Leigh
\thanks{Department of Physics, University of Illinois at Urbana-Champaign,
Urbana IL 61801},
Matthew J. Strassler
\thanks{School of Natural Sciences, Institute for Advanced Study, Olden
Lane, Princeton NJ 08540}
}

\maketitle

\begin{abstract}
Four-dimensional \none\ supersymmetric gauge theories with two
adjoints and a quartic superpotential are believed, from AdS/CFT
duality, to have \slz\ invariance.  In this note we review an old,
unpublished argument for this property, based solely on field theory.
The technique involves a complexified flavor rotation which deforms an
\ntwo\ supersymmetric gauge theory with matter to an \none\ theory,
leaving all holomorphic invariants unchanged.  We apply this to the
\none\ gauge theory with two massless adjoints and show that it has
the same auxiliary torus as that of \nfour\ gauge theory, from which
\slz\ invariance follows.  In an appendix, we check that our arguments
are consistent with earlier work on the $SU(2)$ case.  Our technique
is general and applies to many other \none\ theories.
\end{abstract}

\newpage
\pagestyle{plain}
\narrowtext

\setcounter{footnote}{0}

Recently, the theory of four-dimensional \none\ supersymmetric gauge
theory with two chiral superfields in the adjoint representation has
received considerable attention.  In particular, its \slz\
electromagnetic self-duality and its $AdS_5$ dual representation have
been discussed in numerous papers \cite{twoadjrefs}.  Similar theories
were considered in \cite{klebwit}.  For large gauge groups, \slz\
invariance for this theory has been strongly suggested using its
conjectured duality with type IIB supergravity, which has a
semiclassical symmetry of this type.  Since IIB string theory has
\slz\ as well, then, if one accepts the $AdS$ duality for arbitrary
gauge and 't Hooft couplings, the \slz\ duality of the field theory
should extend to any gauge group.

The \slz\ duality has never been established in the literature from
purely field theoretic considerations.  However, there is an
unpublished argument in favor of this symmetry, and in view of recent
interest in this theory, it seems appropriate to make it more widely
known.  Specifically, we want to consider, for any gauge group $G$,
the \none\ \susic\ gauge theory with two chiral superfields
$\phi_1,\phi_2$ in the adjoint representation and a superpotential
\bel{nthreesup} W = h \ \tr [\phi_1,\phi_2][\phi_1,\phi_2] \ .  \ee
The theory is non-renormalizable and must be defined with a cutoff.
However we are interested in its infrared behavior, where it flows to
a conformal field theory.  The coupling $h$, although canonically of
inverse mass dimension, becomes dimensionless in the infrared.  Define
$h_*$ to be its value in the infrared, in some suitable scheme.  In
\cite{emop} it was shown that there is a continuous set of conformal
field theories near $h_*=0$.  In other words, the quartic
superpotential above is an exactly marginal perturbation of the
interacting conformal field theory with zero superpotential.  The
marginal coupling $h_*$ is inherited from the marginal gauge coupling
$\tau$ of \nfour\ \susic\ Yang-Mills theory, and in \cite{emop} it
was conjectured that the \slz\ self-duality that acts on $\tau$ is
also inherited by the theory with two adjoints.  

In this paper we claim the following.  First, we verify that there
exists a space of conformal field theories which we label by the
parameter $\rho$, or equivalently $q=e^{2\pi i \rho}$, of which $h_*$
is a nontrivial and presumably scheme-dependent function.  Second, we
claim that the set of theories parameterized by $q$, or equivalently
$h_*(\rho)$, has \slz\ duality; the conformal theory with parameter
$\rho$, and the theories with parameter $(a\rho+b)/(c\rho+d)$,
$a,b,c,d$ integers, $ad-bc=1$, are actually different descriptions of
the same theory.  In particular, as seen in the case of $SU(2)$
studied in \cite{kinstwo}, the theory with $h_*=0$ is equivalent to a
theory with a particular non-zero value of $h_*$.

To prove this we employ a complexified flavor symmetry transformation,
under which all holomorphic quantities are invariant even though the
theory as a whole is altered.  One subtlety, not resolved here, is how
precisely to match the dimensionless coupling $\tau$ of the \nfour\
theory on to the coupling $h_*(\rho)$ of the two-adjoint theory.  This
issue need not be settled for the \slz\ invariance to be established.

Let us begin by reviewing the arguments of \cite{emop} concerning the
theory with two adjoints.  Consider an \none\ \susic\ field theory in
four dimensions, with a gauge group $G$ and $N_f$ chiral multiplets
$\phi_i$, $i=1,\dots,N_f$ in the adjoint of $G$.  If $N_f=0$ the
theory is \none\ Yang-Mills and shows confinement and chiral symmetry
breaking.  If $N_f=1$ and the superpotential is zero, the theory is
\ntwo\ supersymmetric and has a Coulomb branch with special points
where magnetically charged BPS states become massless \cite{swone}.
If $N_f=3$ and the superpotential vanishes then the theory is infrared
free.  However, with a renormalizable superpotential
$$
y \tr [\phi_1,\phi_2]\phi_3 
$$ 
the theory will flow to a non-trivial fixed point, becoming
\nfour\ supersymmetric in the far infrared.  If $y=\sqrt 2$, then the
theory is strictly \nfour\ supersymmetric and is conformal at all
scales.  There is a one-complex-dimensional space of such theories,
indexed by the gauge coupling and theta angle through the exactly
marginal parameter $\tau=\theta/2\pi + 4\pi i/g^2$.   

For $N_f=2$, a simple argument shows that with the
non-renormalizable superpotential 
$$
W = h \tr [\phi_1,\phi_2][\phi_1,\phi_2] 
$$
the theory is expected to flow to a point on a one-complex-dimensional
space of conformal fixed points. Specifically, the requirement that
the beta functions for $h$ and for the gauge coupling $g$ both vanish
reduces to a single condition on the two couplings
\cite{emop}.  To see this, note that the anomalous mass dimensions of
$\phi_1$ and $\phi_2$ are equal by symmetry, 
so the beta functions take the form
\bel{betagh}
\beta_g = -f(g)[C_2(G)+2C_2(G)\gamma_\phi]\ ; \
\beta_{h} = h[1+2\gamma_\phi] \ ,
\ee
where these formulas are exact as a consequence of \none\
non-renormalization theorems.  (Here $C_2(G)$ is the second Casimir of
the adjoint representation.)  Only one condition, namely
$\gamma_\phi(g,h) = -1/2$, is required for the two beta functions to
vanish; therefore, if there are any solutions to this condition, they
will typically form one-complex-dimensional subspaces in the
two-complex-dimensional space of couplings.\footnote{If $G$ has a
three-index symmetric invariant, we may consider for $N_f=3$ the
superpotential $ y \tr [\phi_1,\phi_2]\phi_3 + y^s_{ijk} \tr
\{\phi_i,\phi_j\}\phi_k $.  If $y^s\neq 0$, the theory may still be
conformal \cite{finite} but has only \none\ \susy; the space of
conformal theories is three-complex-dimensional (including the \nfour\
subspace.)  Similarly one may add $ h_2 \tr
\{\phi_1,\phi_2\}\{\phi_1,\phi_2\} + h_3[\tr
\{\phi_1,\phi_1\}\{\phi_1,\phi_1\} + \tr
\{\phi_2,\phi_2\}\{\phi_2,\phi_2\}]$ to the superpotential of the
theory with $N_f=2$.  Since the new $\beta$ functions $ \beta_{h_2} =
h_2[1+2\gamma_\phi]; \beta_{h_3} = h_3[1+2\gamma_\phi] $ are
proportional to the other two, the vanishing of all four beta
functions gives only one condition on four couplings,
$\gamma_\phi(g,h,h_2,h_3) = -1/2$, so again we find a
three-complex-dimensional space of fixed points.  More details are
found in \cite{emop}.  }

Unfortunately, the condition $\gamma_\phi(g,h) = -1/2$ can
only be satisfied well outside the realm of perturbation theory, and
it cannot be proven without a shadow of a doubt that solutions exist.
However, there are strong reasons to believe that they are present.
(Recent large-$N$ results \cite{twoadjrefs} support this point of view,
of course.) We will assume for the remainder of this paper that there
is a unique and connected space of solutions to this equation, and that
this space contains a single point with $h=0$ and $g$ equal to 
some special value $g_*$.

We will now focus our attention on the space of conformal theories
with $\gamma_\phi(g,h) = -1/2$. This one-complex-dimensional space has
a single marginal coupling as its parameter. These theories are
particularly interesting as they can be reached through a simple
deformation of \nfour\ Yang-Mills.  In particular, consider \nfour\
Yang-Mills, with gauge coupling $\tau$, deformed by a mass term.
$$
W = \sqrt{2}\ \tr [\phi_1,\phi_2]\phi_3 + \half m_3 \ \tr \phi_3^2
$$
(In the following we will be careful to distinguish $\tau$, the
coupling of the high-energy \nfour\ theory, from both the running
gauge coupling $g$ and the exactly marginal parameter of the
low-energy \none\ conformal theories, which we will call $\rho$.)  At scales
below $m_3$ (more precisely, below some physical scale related
nontrivially to the holomorphic parameter $m_3$) we may integrate out
$\phi_3$.  The theory becomes
$$
W =- {1\over m_3} \tr [\phi_1,\phi_2] [\phi_1,\phi_2] \ .
$$
At the cross-over scale $m_3$ the theory is usefully parameterized by
the gauge coupling $g = 1/\sqrt{{\rm Im}\ \tau}$ and $h\sim -1/m_3$.  At
low energy it flows to a fixed point with couplings $g_*(\tau)$ and
$h_*(\tau)$; the initial gauge coupling of the \nfour\ theory
specifies which \none\ conformal field theory will be reached in the
infrared.  Note that for $\tau\to i\infty$, $h_*\to 0$ but $g_*$ is
finite.

There is a subtlety involved in matching the theory above the scale
$m_3$ with that below $m_3$.  The symmetries of the theory permit
non-perturbative corrections to the superpotential, making it of the
form
\bel{heffinsup} W = h_{eff}\ \tr [\phi_1,\phi_2] [\phi_1,\phi_2] \ ,
\ee where
$$
h_{eff} = -f(q)/m_3, \ f(q) = 1 + \OO (q) \ .
$$
The behavior of $f(q)$ for small $q$ (large ${\rm Im}\,\tau$), the
weak coupling region, is determined by perturbation theory.  The only
direct constraint on $f$ is that it be single-valued under shifts of
the theta angle by $2\pi$, that is $q\to e^{2\pi i}q$.  Higher order
corrections to $f(q)$ are associated with instantons, and have not
been determined beyond low orders.  A similar ambiguity arises if we
attempt to exchange the two parameters $(\tau, m_3)$ for
$(h_{eff},\Lambda)$, where $\Lambda$ is the holomorphic dynamical
scale of the low-energy asymptotically-free gauge theory. By symmetry,
\bel{Lredef} \Lambda^{C_2(G)}\sim
m_3^{C_2(G)}\,g(q), \ g(q) = q + \OO (q^2) \ .  
\ee 
Again, only the leading behavior of $g(q)$ at weak coupling is
determined, and consequently we will not use $\Lambda$ in most of our
discussion.  These ambiguities will not affect our general arguments.

The \nfour\ theory has a duality symmetry, namely an \slz\ symmetry
generated by the semiclassical symmetry $\tau\to\tau+1$ and the
strong-weak coupling transformation $\tau\to -1/\tau$.  This
means the space of \nfour\ theories is smaller than it appears due to
discrete identifications.  Since it seems that there is a one-to-one
map between the space of \nfour\ theories and the infrared two-adjoint
conformal field theories, it is natural to guess \cite{emop} that the
space of two-adjoint conformal theories is also reduced by the same
discrete identifications, taking the form of \slz\ transformations on
the low-energy marginal coupling constant $\rho$. We will call this the
``inherited duality'' conjecture.

We now prove this conjecture, using the following trick.
First, consider the \nfour\ theory with {\it two} mass deformations.
$$
W = \sqrt{2} \tr [\phi_1,\phi_2]\phi_3 + \half m_2\ \tr \phi_2^2
+ \half m_3\ \tr \phi_3^2
$$
If $m_2=-m_3=\hat m$ the theory is \ntwo\ supersymmetric; it is the
theory of \ntwo\ Yang-Mills with a massive adjoint
hypermultiplet, first studied in
\cite{swtwo,nfourtwo}.  The moduli space is a Coulomb branch with an
auxiliary Seiberg-Witten torus and Seiberg-Witten form, as in
\cite{swone}; the torus can be used to specify the low-energy gauge
couplings, and together with the form gives the low-energy K\"ahler
potential (the effective Lagrangian up to second order in momentum.)
The torus is a function of the holomorphic quantities $\hat m^2=
-m_2m_3$ and $q$, and of the holomorphic coordinates on the moduli
space; the low-energy gauge coupling is also a holomorphic
function of these quantities.  For $\hat m=0$ the theory is \nfour\
\susic\ and the auxiliary torus is invariant under \slz, as expected;
for $\hat m\neq 0$ the torus is \slz\ {\it covariant}, although it
can be written in an \slz\ invariant form with a suitable choice
of coordinates on the moduli space \cite{swtwo,nfourtwo}.  

Now consider the global symmetries of this model. For $\hat m=0$ there
is an $SO(6)$ R-symmetry containing an $SU(3)$ flavor symmetry under
which $\phi_i$ are triplets.  Let us consider its $SU(2)$ subgroup
which acts on $\phi_2,\phi_3$ only. Mass terms $m^{ij}\phi_i\phi_j$,
$i,j=2,3$ break the symmetry; $m^{ij}$ transforms in the ${\bf 3}$ of
$SU(2)$. The \ntwo\ case just discussed requires $m^{ij}=
\hat m(\sigma^3)^{ij}$, where $\sigma^a$ are the Pauli matrices.  A 
transformation
$$
\phi_2\to e^{i\alpha}\phi_2 \ , \ \phi_3\to e^{-i\alpha}\phi_3 \ ,
$$
{\it where $\alpha$ is real}, is an $SU(2)$ flavor-symmetry
transformation which changes the phases of $m_2,m_3$ but leaves the
theory invariant.

The observation which permits a proof of the inheritance conjecture
involves the fact that holomorphic quantities, such as the low-energy
gauge coupling, can only depend on the holomorphic product $\det m =
m_2m_3$.  Nothing holomorphic depends on $m_2/m_3$.
Consider the behavior of the theory under the transformation
\bel{suiifrot}
\phi_2\to e^{\alpha}\phi_2 \ , \ \phi_3\to e^{-\alpha}\phi_3\  , 
\ee
{\it where again $\alpha$ is real}.  This imaginary $SU(2)$
transformation is not a symmetry.  \ntwo\ \susy\ is broken to \none, and the
K\"ahler potential, which no longer need depend only on holomorphic
quantities, is altered. However, {\it although the theory
as a whole is not invariant under this transformation, all holomorphic
quantities are unchanged.} In particular, $\det m$ is invariant under
this transformation, and so the torus, and the low-energy gauge
couplings, are unaffected.
It follows, therefore, that for $\det m$ non-zero the low-energy
theory inherits the torus of the \ntwo\ theory with $m_2=-m_3 =
\sqrt{-\det m}$.  

To obtain the theory with two massless adjoints and a quartic
superpotential, we take $m_3\neq 0$ but $m_2m_3=0$.  The torus of this
theory is the same as for $m_2=m_3=0$, the \nfour\ case; it is
invariant under \slz.  At all points on the moduli space where the
low-energy gauge group is abelian, all infrared gauge couplings
$\tau_L$ are equal to the ultraviolet coupling $\tau$.  At the origin
of moduli space, the torus has no massive parameters, indicating a set
of conformal theories indexed by $\tau$, with discrete identifications
under \slz.  This indicates a one-to-one map between the marginal
parameter $\rho$ and the ultraviolet coupling $\tau$, so we may take
$\rho=\tau$. Thus,  the set of conformal field theories with
group $G$ and two adjoint chiral multiplets inherits \slz\
duality from its \nfour\ parent.

It may seem strange at first that the deformation by a mass term
$\half m_3\tr\phi_3^2$ does not change the low-energy torus and its
attendant gauge couplings.  However, there is a simple physical
explanation, most easily presented in the case of $G=SU(2)$.  In this
case the theory has a moduli space with a single coordinate $u=\half
\tr(\phi_1^2)$.  For given $\vev{u}$, the gauge group is broken to
$U(1)$ and the low-energy gauge coupling $\tau_L$ is the modular
parameter of a holomorphic torus \cite{swtwo}, reproduced in
\eref{swtorus}, which is a function of $\hat m^2$, $u$, and $\tau$.
The imaginary global symmetry transformation \eref{suiifrot} breaks
the supersymmetry to \none, but the gauge coupling is still the
modular parameter of a holomorphic curve \cite{kinsone}.  Since the
complexified flavor rotation \eref{suiifrot} leaves $\det m$, $u$ and
$\tau$ invariant, this curve is the same as that of the \ntwo\ theory
with a massive adjoint hypermultiplet, except that ${\hat m}^2$
is replaced with $-m_2m_3$ in \eref{swtorus}. But how
can it make physical sense that the \loe\ \uone\ coupling $\tau_L$
should not depend in any way on $m_2/m_3$?  This is quite easy to see
in the weak-coupling limit.  Let us take the expectation value of $u$
to be large compared with $m_2$ and $m_3$, so that perturbation theory
is valid.  The \suii\ group is broken to \uone\ at the scale
$\sqrt{u}$; the fields $\phi_2$, $\phi_3$ are massive charged fields,
with mass matrix 
\bel{massmatr}\left[
\begin{array}{cc} m_2 & 2\sqrt{u} \\ 
-2\sqrt{u} & m_3 \end{array} \right] \ .  
\ee
Referring to their masses as $\mu_2$ and $\mu_3$, we note
$\mu_2\mu_3=4 u+m_2m_3$.  Consider the case where
$|\mu_3|>|2\sqrt{u}|>|\mu_2|$. (All other cases lead to the same
result, though the description of the physics will be different.)
Above both masses, the coupling does not run.  Between $\mu_3$ and
$2\sqrt{u}$, the \suii\ theory has only two adjoints and runs toward
strong coupling, generating a logarithm of $\mu_3/2\sqrt{u}$ with
beta-function coefficient $-2$.  At the scale $2\sqrt{u}$ the gauge
group is broken to \uone\, and the remaining charged fields cause the
coupling to run toward weak coupling, generating a logarithm of
$2\sqrt{u}/\mu_2$ with beta-function coefficient $+2$.  Below the
scale $\mu_2$ the coupling ceases to run, and the \loe\ coupling
constant is
\bel{taurun} \tau_L=
\tau+ {1\over i\pi}\log\left[{\mu_2\mu_3\over 4u}\right] \ , 
\ee
which depends only on $u$ and on $m_2m_3$.  Evidently, the effects of
raising $m_3$ and lowering $m_2$ are being arranged to cancel.
Remarkably, this perturbative cancellation generalizes to the full
non-perturbative behavior of the holomorphic properties of the theory.

So far we have learned that varying some massive parameters in the
theory of \suii\ with three adjoints can change the theory at non-zero
momenta (influencing the \kpot\ and massive states) while leaving the
holomorphic part of the far-infrared physics the same.  In particular,
\slz\ invariance is preserved.  But now we may consider the limit
where $|m_3|\gg |u|,|m_2|$.  Below $m_3$ we have the theory with
superpotential \eref{nthreesup}, along with a mass for $\phi_2$.  For
small $q$ and $|qm_3^2|\ll |u|\ll |m_3^2|$, the gauge coupling is weak
everywhere and the analysis involving Eqs.~(\ref{massmatr}) and
(\ref{taurun}) still applies.  In the limit $m_2\ll u/m_3$, we have
$\mu_3\approx m_3$, $\mu_2\approx 4u/m_3$, with $\mu_2\mu_3=4 u$; the
coupling runs with a beta function of $-2$ between $m_3$ and
$2\sqrt{u}$, then with beta function $+2$ between $2\sqrt{u}$ and
$2u/m_3$.  Thus, in the $m_2\to 0$ limit, all running effects above
and below $2\sqrt{|u|}$ cancel precisely in \eref{taurun}, and we find
that the gauge coupling $\tau_L$ of the low-energy \uone\ theory is
the same as that of the high-energy theory, $\tau$, just as in the
unbroken \nfour\ gauge theory.  This reflects the fact that the torus
depends in this limit only on $q$ and $u$, and not on $m_3$.
(Another example of this type appears in \cite{gremmkap}.)

As in the \nfour\ theory, the limit $u\to 0$ leaves the torus a
scale-invariant \slz-invariant function of the parameter $q=e^{2\pi i
\tau}$. We take this as evidence that at $u=m_2=0$ the theory
\eref{nthreesup} flows to a non-trivial conformally invariant theory
in the infrared.  Indeed there is a set of conformal field theories
indexed by $q$, in agreement with \cite{emop} and \eref{betagh}.

The full picture for the $SU(2)$ case is now the following.  The
\nfour\ conformal theories are parameterized by $\tau$, or
equivalently $q=e^{2\pi i \tau}$ living on the disk $|q|\leq 1$.
These theories are identified under \slz\ transformations.  In the
\none\ conformal theories with two adjoints and superpotential
\eref{nthreesup}, there is also a marginal parameter $\rho$, which we
may choose to be numerically equal to $\tau$, such that $q = e^{2\pi i
\rho}$ has the same properties as $q$ in the \nfour\ theory.  The
limit $q=0$ corresponds to $h=0$, but the general relation between $h$
and $q$ is nontrivial and presumably scheme-dependent. The
transformations $\rho \to -{1\over \rho}$ and $\rho\to \rho +1$
correspond to electric-magnetic and magnetic-dyonic dualities.  In the
appendix we show, using the integrating-in techniques of \cite{intin},
that these claims agree with those described in \cite{kinstwo}.

There are many other interesting \none\ theories which will also have
large discrete invariance groups.  The simplest can be generated by
taking the theory of \ntwo\ \suii\ with four \hyp s.  This theory was
shown by Seiberg and Witten to have \slz\ invariance as well
\cite{swtwo}.  The flavor symmetry of the doublets is $SU(8)$, which
is broken by the \ntwo\ superpotential term to $SO(8)$.  By giving
masses to some of the quarks and performing complexified $SO(8)$
transformations, we can again find many theories with
non-renormalizable operators which are exactly marginal in the
infrared, and whose couplings transform under \slz.
Examples of such theories appear in \cite{newnone,kapold}.

Another related theory is \ntwo\ \susic\ $SU(N)\times SU(N)$ with two
hypermultiplets in the $({\bf N,\bar N})$ representation.  This has a
duality group given by the symmetry group of a torus with two
identical marked points \cite{ewelliptic}.  The superpotential $W=
\Phi_1(Q_1\tilde Q_1+Q_2\tilde Q_2)+ \Phi_2(\tilde Q_1Q_1+\tilde Q_2
Q_2) $ has a $U(2)$ flavor symmetry, broken by masses $
m^{ij}Q_i\tilde Q_j$, under which $m$ is a triplet plus a singlet.  If
equal and opposite masses for the hypermultiplets are added, a
complexified flavor rotation can leave the theory with mass terms
$mQ_1\tilde Q_2 + m'Q_2\tilde Q_1$.  With $m'$ finite and $m=0$ we
obtain a set of conformal theories, with a quartic superpotential
involving $\Phi_1,\Phi_2,Q_1,\tilde Q_2$, that inherits the duality
group of its parent.

Our technique is a simple and powerful tool for showing that large
duality groups are widespread in \none\ supersymmetry.  It would be
interesting to understand more deeply the mathematical underpinning
of these results, and to find a brane-based realization of
our method.

\acknowledgements

We would like to thank N. Dorey, N. Seiberg and E. Witten for discussions.
K.I. and M.J.S. were supported by National Science Foundation grant
NSF PHY95-13835 and by the W.M. Keck Foundation.  K.I. was also
supported by UCSD grant DOE-FG03-97ER40546. P.C.A. was supported by
National Science Foundation grants NSF PHY94-07194 and NSF PHY95-13717
and by an A.P. Sloan Foundation fellowship.  This work was done mostly
at Rutgers University, during which time M.J.S., P.C.A. and R.G.L were
supported by Department of Energy contract DE-FG05-90ER40559.
   
\section*{Appendix}

In this appendix we show that the application of our techniques to the
$SU(2)$ theory with two adjoints reproduces the results of
\cite{kinstwo} for $SO(3)$ with two triplets.

Consider $SU(2)$ ${\cal N}=4$ Yang-Mills with gauge coupling $\tau$
deformed by ${\cal N}=1$--preserving masses, giving the superpotential
$$
W= \sqrt{2} \tr \phi_1[\phi_2,\phi_3] + m^{ij}u_{ij}
$$
where classically $u_{ij}=\half\tr(\phi_i\phi_j)$.
Denote the three eigenvalues of $m^{ij}$ by $m_i$, and define 
$u\equiv u_{11}$.

Our complexified flavor rotation trick implies that for $m_1=0$, the
low-energy effective coupling $\tau_L$ on the Coulomb branch of this
theory is equal to that of the $SU(2)$ ${\cal N}=2$ theory with a
fundamental hypermultiplet of mass ${\hat m}^2 = -m_2 m_3$.  In
\cite{swtwo}, $\tau_L$ is given as the modular parameter of the
auxiliary torus
\bel{swtorus} 
y^2 = \prod_{i=1}^3 \left(x-e_i(q)\tilde u + {1\over4}
e_i^2(q) m_2 m_3 \right) \ ,
\ee 
where
\bel{utilde}
\tilde u \equiv \vev{u} + (1/8)e_1(q) m_2 m_3 ,
\ee
$q\equiv e^{2\pi i\tau}$, and the
$e_i(q)$ are the usual modular forms associated with the torus,
satisfying $e_1+e_2+e_3=0$, $e_1-e_2=[\theta_3(\tau)]^4$, {\it
etc.}, with small-$q$ expansions
\begin{eqnarray}\label{eexp}
e_1(q) &=& {2\over3} + 16q + {\cal O}(q^2),\nonumber\\
e_2(q) &=& -{1\over3} - 8q^{1/2} - 8q + {\cal O}(q^{3/2}),\nonumber\\
e_3(q) &=& -{1\over3} + 8q^{1/2} - 8q + {\cal O}(q^{3/2}).
\end{eqnarray}
For fixed $m_2m_3$ and $\tilde u$ the torus is \slz\ invariant.  This
follows from the modular properties of the $e_i$, which are
interchanged with one another under \slz.  In particular, $e_1
\leftrightarrow e_2$ under $\tau \to -1/\tau$, while $e_2 
\leftrightarrow e_3$ under $\tau \to \tau +1$.  Note that with the
definition of $\tilde u$ given in \eref{utilde}, the Coulomb
branch coordinate $\vev{u}$ transforms under \slz.

There is, however, an ambiguity involved in the definitions of $u$ and
the masses.  The symmetries of the theory permit non-perturbative
redefinitions of the form
\bel{uredef}
u \to k_1(q) u + q\, k_2(q) m_2 m_3, \qquad k_i(q) = 1 + {\cal O}(q).
\ee
and
\bel{mredef}
m_i \to \ell_i(q) m_i, \qquad \ell_i(q) = 1 + {\cal O}(q).
\ee
The reason is that $u$ and the bare masses are only defined in the
weak coupling ($q\to0$) limit, which the above redefinitions preserve.
(More generally, $u_{ij} \approx \half \tr (\phi_i  \phi_j)$ and
$m^{ij}$ can suffer such redefinitions, with $u_{ij}$ mixing at order
$q$ with the subdeterminants of $m^{ij}$.)

It is convenient to use the above freedom to redefine the masses by
\bel{newmass}
m_2m_3 \to 9 e_2(q) e_3(q) m_2m_3,
\ee
which can be checked to be of the form of \eref{mredef} using
\eref{eexp}.  The virtue of this redefinition is that, when
used in \eref{utilde}, $\vev{u}$ is \slz\ invariant.  This is
convenient for keeping manifest \slz\ invariance in our calculations
when we integrate $u$ out, as we will do shortly.

We can now recover the results of \cite{kinstwo} for the $SO(3)$
theory with two triplets and no superpotential.  This is the limit of
our theory in which $m_3\to \infty$, $q\to 0$ keeping $\Lambda \sim
m_3\sqrt{q}$ fixed.  In \cite{kinstwo} three descriptions of the
theory were found, in terms of electric, magnetic, and dyonic states.
The electric description has no superpotential; the magnetic (dyonic)
description (upon integrating out some massive singlets used in the
presentation of \cite{kinstwo}) has superpotential
\bel{qsup} 
W = -{ \eta\over 8\Lambda} \det_{j,k} [\tr (\phi_j\phi_k)] 
\ee 
where $\eta=1$ $(-1)$ in the magnetic (dyonic) description.

Now, the $SU(2)$ theory with one massless and two massive adjoints
has a superpotential which enforces the relation (\ref{swtorus}),
with the substitution of \eref{newmass}, by
way of a Lagrange multiplier $\lambda$
$$
W= \lambda\left[ y^2 - \prod_{r=1}^3
\left(x - e_r\left[ u + {9\over 8}e_1e_2e_3 m_2 m_3\right]
+{1\over 4} e_r^2 m_2 m_3\right)\right] .
$$
Adding a mass for $\phi_1$ takes $W\to W + m_1 u$.
Upon integrating $u$ out, one finds 
a low energy superpotential with three branches
\bel{intoutsup}
W_{r,L} = m_1 u_r(q,\hat m)|_{{dW\over du}=0} =  -{1\over8}m_1m_2m_3 
[ 9 e_1(q)e_2(q)e_3(q) + 2 e_r(q) ]
\ee
where $u_r$ is one of the three values of $u$ at which the torus
(\ref{swtorus}) becomes singular.  (See also \cite{dorey}.)

Taking $m_1m_2 = \det\widetilde m$ where $\widetilde m^{jk}$ for
$j,k=1,2$ is the mass matrix for $\phi_{1,2}$, and integrating the two
adjoint fields $\phi_1$ and $\phi_2$ back in, as in \cite{intin}, gives
$$
W_r = \left[W_{r,L} -  \widetilde m^{jk} u_{jk}
\right]_{{dW\over d\tilde m}=0}
= {2 \over m_3[9 e_1e_2e_3 + 2 e_r]} \det_{j,k=1,2} u_{jk}
$$
The $q\to 0$, $m_3 \to \infty$ limit, for which
$u_{ij}\to\half\tr(\phi_i\phi_j)$,  gives
\begin{eqnarray}
W_1 &=& {4 \det_{j,k} u_{jk}\over m_3 e_1 (9e_2e_3+2)} 
\to 0, \nonumber\\
W_2 &=& {4 {\det_{j,k} u_{jk}}\over m_3 e_2 (9e_1e_3+2)}
\to -{\det_{j,k} [\tr (\phi_j\phi_k)]\over 8\Lambda}, \nonumber\\
W_3 &=& {4 {\det_{j,k} u_{jk}}\over m_3 e_3 (9e_1e_2+2)} 
\to {\det_{j,k} [\tr (\phi_j\phi_k)]\over 8\Lambda}, \nonumber
\end{eqnarray}
matching to the electric, magnetic, and dyonic superpotentials of
\eref{qsup}, respectively.  In this way we see explicity how the \slz\
duality transformations $\tau \to -{1\over \tau}$ and $\tau \to \tau
+1$ correspond to electric-magnetic duality and magnetic-dyonic
duality.  A $\Gamma_2$ subgroup of \slz\ leaves the descriptions
invariant, while $SL(2,{\bf Z})/\Gamma_2 \cong S_3$ permutes the three
descriptions.  These facts were already understood in \cite{kinstwo}
from the connection of the theory with the duality of \nfour\ \suii\
gauge theory; here we understand it as following from the \slz\
duality of the \none\ theory itself.  Furthermore, it is easy to check
that redefining the fields and parameters by putting in arbitrary
$g(q)$, $k_i(q)$, and $\ell_i(q)$ [see Eqs.~(\ref{Lredef}),
(\ref{uredef}), and (\ref{mredef})] has no effect on the above
calculation.

One could also attempt to start from the low energy superpotentials in
\eref{intoutsup} and integrate back in all three $\phi_j$; we then
expect to recover the ${\cal N}=4$ theory with coupling $q$.  However,
this is more difficult than in the above two-flavor case because we
must introduce not only $\tr(\phi_i\phi_j)$ but also the the gauge
invariant non-quadratic operator $\det (\phi)$.  (Here $\phi$ is a
$3\times 3$ matrix in flavor and color.)  So we instead consider going
in the other direction, attempting to recover \eref{qsup} in the
theory with two adjoints by integrating massive $\phi_3$ out of the
${\cal N}=4$ theory.

Consider the theory with three adjoints, superpotential
$W=\sqrt{2}\beta \det \phi$, and coupling $q$.  For $\beta =1$ the
theory is conformal.  For $\beta \neq 1$ the theory has only
\none\ supersymmetry, but flows until it
reaches the IR attractive ${\cal N}=4$ fixed point where the physical
(nonholomorphic) coupling $\beta =1$.
As discussed in \cite{kinstwo}, the quantity $t\equiv \beta ^4 q$ is invariant
under this RG flow (more generally, $\beta ^{2C_2(G)}q$ is invariant),
and the low-energy \nfour\ conformal theory has coupling $t$.
The theory is nowhere weakly coupled unless $t\ll 1$.

Adding ${1\over 2} m_3 u_{33}$ to the superpotential and integrating
out $\phi_3$, we find that symmetries ensure that
the low-energy superpotential is
\bel{sdef}
W_L={\beta ^2 s(t)\over m_3} \det_{i,j=1,2} \tr(\phi_i \phi_j). \ee
where $s(t)$ is an unknown function.  (The symmetries ensure that in
this case $u_{ij}$ and $\half\tr(\phi_i \phi_j)$ are proportional,
differing by another function of $t$.)  For $\beta$ fixed and
$q\rightarrow 0$, the theory is weakly coupled and we may integrate
out $\phi_3$ classically, which reveals that $s(t=0)=1$. At finite
$t$, $s(t)$ is undetermined.  However, we know the $\beta=1,q\to 0$
theory is \slz--dual to $\beta=1, q\to 1$ and $\beta=1, q\to e^{2\pi
i}$, which are the magnetic and dyonic descriptions.  In these
descriptions, where $t\sim 1$, the classical analysis is not valid,
and $s(t)$ may differ from 1.  We now determine $s(1)$ in the scheme
used in \cite{kinstwo}.

Taking $\beta =1$, $q\rightarrow 0$, $m_3\rightarrow \infty$, with
$\Lambda = m_3 q^{1/2}$ held fixed, \eref{sdef} obviously yields the
expected electric low-energy superpotential, \eref{qsup} with $\eta
=0$.  A magnetic description of the same theory should be obtained by
studying the theory with $\beta =1$ and $q\to 1$, but it is convenient
instead to study a theory with the same {\it infrared} physics, namely
one with $\beta=q^{-1/4}$ and $q\to 0$; both theories have $t=1$.  The
latter theory can be defined by holding the strong coupling scale
$\Lambda\equiv m_3q^{1/2}$ fixed, as in \cite{kinstwo}. From \eref{sdef}
this limit has superpotential
$$
W_L= {s(1)\over m_3q^{1/2}}\det_{i,j=1,2}\tr(\phi_i
\phi_j)
$$ 
which agrees with \eref{qsup} for $\eta=1$ provided $s(1)=1/8$.  The
dyonic description is given by taking $q\to e^{2\pi i}$, which changes
the sign of the superpotential in agreement with \eref{qsup}.

\end{document}